\documentclass[10pt,letterpaper,twocolumn]{article} 
\usepackage{ol2}
\usepackage[draft,implicit=false]{hyperref}
\usepackage{amsmath}

\begin{document}

\twocolumn[ 

\title{\emph{Gap solitons with null-scattering}}


\author{K. Nireekshan Reddy and S. Dutta Gupta$^{*}$ }

\address{
School of Physics, University of Hyderabad, Hyderabad-500046, India
\\
$^*$Corresponding author: sdghyderabad@gmail.com
}

\begin{abstract}
We study excitation of gap solitons under the conditions of coherent perfect absorption (CPA). Our system consists of a symmetric periodic structure with alternating Kerr nonlinear and linear layers, illuminated from both the ends. We show near-total transfer of incident light energy into the gap solitons resulting in null-scattering. We also report on the nonlinear super-scattering (SS) states. Both the CPA and the SS states are shown to be characterized by typical field distributions. Both the exact and the approximate results (based on nonlinear characteristic matrix method) are presented, which show good agreement.
\end{abstract}


 ] 
Nonlinear stratified media, in particular, nonlinear periodic media have drawn considerable attention over the past few decades due to their tremendous application potentials\cite{sdg-review}. In recent years there is a renewed interest because of fundamental issues involving $\mathcal{PT}$ symmetry and the associated spectral singularities  \cite{pt-nonlinear1,pt-nonlinear2,pt-gap-1,pt-gap-2,pt-singl}. In the context of illumination from one side, particular attention was given to nonlinear periodic structures where the incident light frequency was chosen in the stop gap. It was shown that nonlinearity induced total transmission through the structure can be effected when the  field distribution inside the layered medium corresponds to soliton-like spatial patterns\cite{chen-mills-prl,chen-mills-prb}. A great deal of work has been carried out on such gap solitons both with continuous  and discrete periodicities \cite{sdg-review,sdg-imperial,sipe-review,molemed}.~There are also reports on such gap solitons in $\mathcal{PT}$ symmetric systems \cite{pt-gap-1,pt-gap-2}. Exact solutions for gap solitons for $s$-polarized light is now known \cite{chen-mills-prb}. Simultaneously approximate methods like nonlinear characterstic matrix method (NCMM) were also developed to capture the same physics \cite{ncmt,sdg-review}. These methods were extensions of the linear characteristic matrix theory based on slowly varying envelope approximation (SVEA) \cite{marburg}.
\par
From a different angle there have been efforts to integrate the notions of recently invented coherent perfect absorption (CPA) \cite{cpa1,cpa-longhi,sdg-opex,sdg-ol12} with nonlinearity  \cite{ncpa2,ncc,ncpa-knr-sdg}. It was shown that nonlinearity can play a vital role in restoring CPA in off-resonant systems \cite{ncpa-knr-sdg,ncc}. In the context of recent advancements in the theory and experiments on CPA, it is an interesting problem to integrate the notions of CPA with gap solitons. One can ask the question if it is possible to channel all the incident plane wave energy into the gap solitons. Clearly this would mean CPA like null-scattering leading to an energy efficient system. In this letter we investigate a symmetric periodic structure with alternating Kerr nonlinear and linear layers. Like in CPA the system is illuminated from both left and right ends \cite{sdg-opex}. We exploit the symmetry of the structure to calculate the incident intensities under which there is little or no outgoing waves from the structure. We also study the super scattering (SS) states which correspond to constructive interference as opposed to CPA. Depending on the parameters of the system, the null-scattering states correspond to one-, two-, and more-, soliton-like intensity distributions in the nonlinear layered medium. Calculations are performed using the NCMM and confirmed by exact numerical integration of the differential equations. We present details of each of the methods. The agreement of the approximate and the exact theory validates the NCMM applied to such nonlinear systems.
\par
\begin{figure}[ht]
\center{\includegraphics[width=8cm]{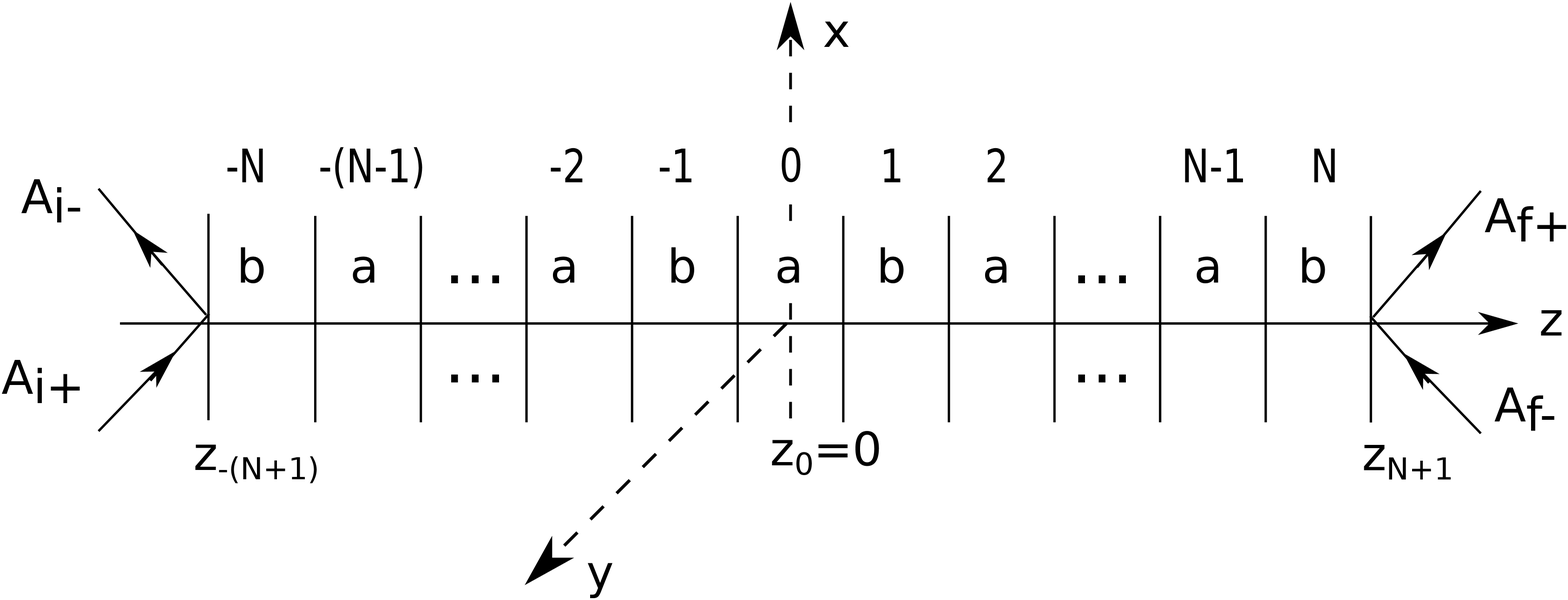}}
\caption{Schematic view of the symmetric periodic layered medium with $N$ periods consisting of `$a$' and `$b$' type layers. `$a$' (`$b$') type layer is nonlinear (linear) with dielectric function $\bar{\epsilon}_a$ ($\epsilon_b$) and width $d_a$ ($d_b$). Keeping symmetry in view we have labeled the layers from the central one.} 
\label{fig:fig1}
\end{figure}
Consider the periodic layered medium shown in Fig.~\ref{fig:fig1}. Let the number of periods be $N$ resulting in $2N+1$ layers for the total structure. We assume the central layer to be Kerr nonlinear and extreme layers to be linear with lower refractive index. The dielectric function of the Kerr nonlinear slab is given by
\begin{equation}
\bar{\epsilon}_a=\epsilon_a+\alpha|E|^2,
\label{eq1}
\end{equation}
where $\alpha$ is the nonlinearity constant. Let the structure be illuminated at an angle, $\theta$ from both the sides by monochromatic $s$-polarized waves of wavelength $\lambda$.
Moving to dimensionless variables as 
\begin{equation}
 z\rightarrow k_0z \hspace{0.1cm}(k_0=\omega/c),\hspace{0.25cm} E\rightarrow \sqrt{\alpha} E, \label{eq2}
\end{equation}
one can write down the Maxwell's equations for tangential components of electric and magnetic fields in any $j^{th}$ layer as
\begin{eqnarray} \label{eq3}
\frac{dE_{jy}}{dz}&=&-iH_{jx},\\ 
\frac{dH_{jx}}{dz} &=&\begin{cases} \label{eq4}
       -i[(\epsilon_j-p_{x}^2)+|E_{jy}|^2]E_{jy} \hspace{0.15cm} \text{for even $j$,}\\ \\
       -i[(\epsilon_j-p_{x}^2)]E_{jy} \hspace{0.15cm} \text{for odd $j$}, 
     \end{cases} 
\end{eqnarray}
where $p_x=\sqrt{\epsilon_{i}}\sin{\theta}$. Note that $E_{jy},~H_{jx}$ are in general complex and one has four coupled nonlinear equations for the real and imaginary parts of the fields. We have numerically integrated this set with proper boundary conditions to obtain the exact results (see below).
\par
In what follows we present the outline of the approximate NCMM which has been used extensively in the past \cite{sdg-imperial,sdg-review,localization}. The propagation through each nonlinear layer is  obtained in terms of the approximate (under SVEA) solution for the electric field as \cite{marburg}
\begin{equation}
E_{jy}=A_{j+}e^{ip_{jz+}(z-z_j)}+A_{j-}e^{-ip_{jz-}(z-z_j)}, \label{eq5}
\end{equation} 
with $z_j \leq z \leq z_{j+1},~z_0=0$ and $p_{jz+}~(p_{jz-})$ denotes the normalized (to $k_0$) $z$ component of propagation constant for forward (backward) propagating waves with amplitudes $A_{j+}~(A_{j-})$. $p_{jz\pm}$ are given by the following
\begin{equation}
p_{jz\pm}=\begin{cases}
\sqrt{\epsilon_j-p_x^2+(|A_{j\pm}|+2|A_{j\mp}|)} \hspace{0.2cm} \text{for even $j$}, \\ \\
\sqrt{\epsilon_j-p_x^2}\hspace{0.2cm} \text{for odd $j$}. 
\end{cases}
\label{eq6}
\end{equation} 
We choose $\theta$ in range 
\begin{equation}
\epsilon_a-p_x^2>0, \hspace{0.5cm} \epsilon_b-p_x^2<0,\label{eq7}
\end{equation}
in order to ensure that the waves are propagating (evanescent) in nonlinear (linear) layers. Thus the structure shown in Fig. \ref{fig:fig1} represents a system of $N$ coupled nonlinear wave guides, sandwiched between two identical high index prisms representing a resonant tunneling geometry \cite{res-tunnel,res-tunnel-sdg,yeh}. Thus transmission through the structure can be realized when the resonant modes (the so called supermodes) are supported by the coupled guides \cite{yeh,sdg-imperial}. For the illumination angle in the stop gap, total transmission of the nonlinear system for one sided illumination is known to lead to gap solitons \cite{sdg-imperial,chen-mills-prb,sipe-review,chen-mills-prl,sdg-review}.
\par 
The column vectors with the tangential field components $E_{jy}$, $-H_{jx}$ as components, can be related at two different planes $z=z_j$ and $z=z_{j+1}$ as
\begin{eqnarray}
\begin{pmatrix}
E_{jy} \\ -H_{jx} \end{pmatrix}_{z=z_{j}} &=&{{M}_j}
\begin{pmatrix}E_{jy} \\-H_{jx} \end{pmatrix}_{z=z_{j+1}}, \label{eq8}
\end{eqnarray}
where $M_j$ denotes the nonlinear (linear) characteristic matrix \cite{sdg-review} when $j$ is even (odd). For CPA like excitation scheme (with both sided illumination) we exploit the symmetry (as in  \cite{ncpa-knr-sdg}) of the structure allowing only the symmetric ($A_{0+}=A_{0-}=A_{0}$) and the antisymmetric ($A_{0+}=-A_{0-}=-A_{0}$) solutions. This results in $p_{0z+}=p_{0z-}=p_{z0}$. By treating $A_0$ as a  parameter one can calculate the incident amplitude $A_{i+}(A_{f-})$ and the scattered amplitudes $A_{i-}(A_{f+})$ on the left (right) sides outside the layered medium.
For example, starting from $z_0=0$ in the central layer one can calculate the amplitudes in the nonlinear layer labeled by $j=-2$ as
\begin{eqnarray}
\begin{pmatrix}
A_{-2+} \\ A_{-2-} \end{pmatrix} = \begin{pmatrix}  1 & 1 \\ p_{-2z+} & -p_{-2z-} \end{pmatrix}^{-1}M_1(d_b)\times \nonumber \\ \times M_0(d_a/2)\begin{pmatrix}  1 & 1 \\ p_{0z} & -p_{0z} \end{pmatrix}\begin{pmatrix}A_0 \\ \pm A_0 \end{pmatrix}, \label{eq9}
\end{eqnarray}
where $M_1(d_a/2)$ and $M_2(d_b)$ denote the characteristic matrices for  nonlinear and linear layers of widths $d_a/2$, $d_b$, respectively. Since one doesn't know $A_{-2\pm}$ \textit{apriori}, $p_{-2\pm}$ are also not known. $p_{-2\pm}$ and hence $A_{-2\pm}$ are then evaluated by taking $|\cdots|^2$ on both the sides of Eq.~(\ref{eq9}) yielding  the corresponding intensity relations and then using fixed point iteration for the coupled nonlinear equations \cite{sdg-imperial}. Continuing the same procedure for all the layers, for example, till the left end we have
\begin{eqnarray}
\begin{pmatrix}
A_{i+} \\ A_{i-} \end{pmatrix} = \begin{pmatrix}  1 & 1 \\ p_{iz} & -p_{iz} \end{pmatrix}^{-1} M_N(d_b) M_{N-1}(d_a) \cdots \nonumber \\ \cdots M_1(d_b)M_0(d_a/2)\begin{pmatrix}  1 & 1 \\ p_{0z} & -p_{0z} \end{pmatrix}\begin{pmatrix}A_0 \\ \pm A_0 \end{pmatrix}, \label{eq10}
\end{eqnarray} 
where $p_{iz}$ denotes the $z$ component of the propagation constant for the incident (scattered) wave of amplitude $A_{i+}$~($A_{i-}$) on the left side of the structure. One can also propagate towards right and evaluate the scattered ($A_{f+}$) and the incident ($A_{f-}$) wave amplitudes. As a consequence of inherent symmetry we find that for symmetric (antisymmetric) solution $A_{i+}=A_{f-}=A_{in}$ ($A_{i+}=-A_{f-}=-A_{in}$) and $A_{i-}=A_{f+}=A_{t}$ ($A_{i-}=-A_{f+}=-A_{t}$)\cite{ncpa-knr-sdg}. We can thus define the normalized scattering intensity as $S=|A_{i-}/A_{i+}|^2=|A_{f+}/A_{f-}|^2=|A_{t}/A_{in}|^2$ as a quantitative measure of CPA. 
\par
We now outline the symmetry considerations for the exact numerical integration. With the origin as defined in Fig.~\ref{fig:fig1}, the supported solutions are either symmetric or antisymmetric with
\begin{eqnarray}
E_y(-z)&=E_y(z),\hspace{0.5cm}H_x(-z)=-H_x(z),&\\ \label{eq11}
E_y(-z)&=-E_y(z),\hspace{0.5cm}H_x(-z)=H_x(z),&  \label{eq12}
\end{eqnarray}
respectively. Thus at $z=0$ in the central layer we have $H_x=0$ ($E_y=0$) for the symmetric (antisymmetric) solution. The other non vanishing component $E_y$ ($H_x$) at $z=0$ is treated as a parameter for symmetric (antisymmetric) solution. The initial conditions at $z=0$ for the tangential components of the electric and magnetic fields are thus known and Eqs.~(\ref{eq3})-(\ref{eq4}) can be numerically integrated till the left/right end ensuring the continuity of the fields at every interface. Since the prisms are assumed to be linear the incident and scattered intensities can be computed in a straight forward manner from the knowledge of tangential fields.
\begin{figure}[pb]
\center{\includegraphics[width=8cm]{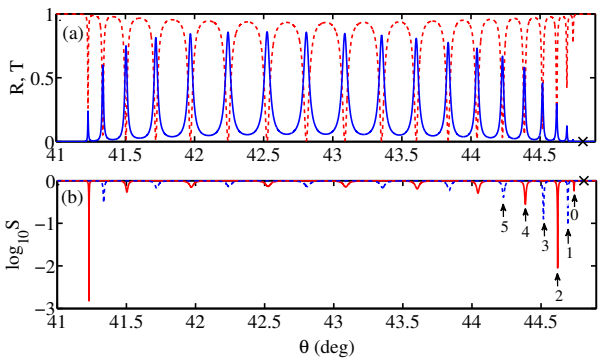}}
\caption{Linear responses of the system: (a) Reflection $R$ (dashed) and transmission $T$ (solid) coefficients and  as functions of $\theta$ for unidirectional illumination; (b) Intensity scattering $\log_{10}S$ vs. $\theta$ for bidirectional illumination for symmetric (solid) and antisymmetric (dashed) solutions. The CPA-like dips are labeled from the extreme right with even (odd) integers for symmetric (antisymmetric) cases.} 
\label{fig:fig2}
\end{figure}
\par
We now present our numerical results. Most of the results presented are exact, obtained by numerical integration of Eqs.~(\ref{eq3})-(\ref{eq4}). Direct numerical integration is used instead of the known exact solutions for Kerr nonlinear layers for $s$-polarized light\cite{chen-mills-prb} with future goals in mind. The technique used here is valid for any form of nonlinearity including the saturation type, which is attracting a lot of attention in the context of regularization of spectral singularities \cite{liu-sdg-gsa}. We also included some NCMM results to compare with exact ones to check the validity of the SVEA, which is often used in many other nonlinear systems. Throughout our calculations we have taken $\epsilon_a=4.0$,~$\epsilon_b=2.25+i~0.5\times10^{-3}$,~$N=19$,~$\epsilon_i=6.145$,~$d_a=0.2\lambda$,~$d_b=0.365\lambda$ and $\theta$ is varied in between $41.0^{\circ}$ and $45.0^{\circ}$, where the condition given by Eq.~(\ref{eq7}) is satisfied.
\par
We first study the linear properties. Let the amplitude reflection and transmission coefficients with unidirectional illumination be denoted by $r$ and $t$, respectively. The results for the corresponding intensity reflection $R=|r|^2$ and intensity transmission $T=|t|^2$  as functions of the angle of incidence are shown in Fig.~\ref{fig:fig2}(a). The sharp resonances in Fig.~\ref{fig:fig2}(a) in a small range of angles are due to resonant tunneling of electromagnetic radiation. The number of such resonances coincides with the number of the guides as expected. At every resonance we have the characteristic phase jump of $\pi$ (not shown here) between reflected and transmitted light ($\Delta\phi=|\phi_r-\phi_t|$).~Thus if one satisfies $|r|=|t|$ for unidirectional illumination near any of these resonances then one can have CPA for two sided illumination geometry \cite{sdg-opex}.~We studied the linear CPA for bidirectional illumination of the structure for both symmetric and antisymmetric solutions.~The CPA-like dips in scattering are shown in Fig.~\ref{fig:fig2}(b) for given losses and system parameters.~We label these dips by integers $0,~2,~4$ ($1,~3,~5$) for symmetric (antisymmetric) solutions from the extreme right end. A prominent CPA-like dip can be seen in Fig.~\ref{fig:fig2}(b) for the resonance labeled by $2$ as $|r|\approx|t|$ at that particular angle (see Fig.~\ref{fig:fig2}(a)) and the phase condition is met by characteristic phase jump at the resonance.
\begin{figure}[]
\center{\includegraphics[width=8cm]{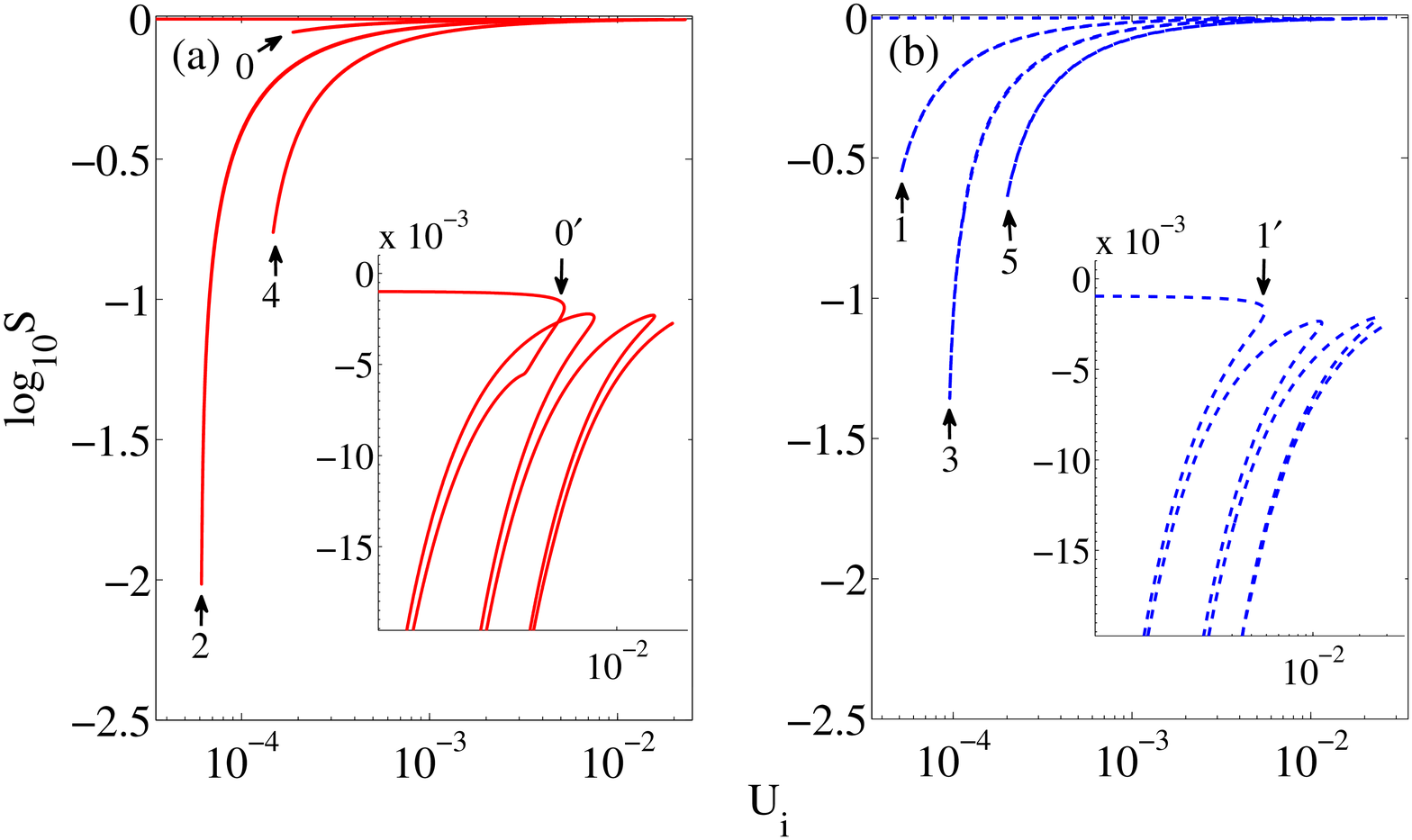}}
\caption{ Nonlinear response of the structure for bidirectional illumination. Intensity scattering ($\log_{10}S$) as a function of incident intensity $U_i$ in the stop gap with $\theta=44.81^{\circ}$ 
for (a) symmetric and (b) antisymmetric solutions. The CPA-like dips are labeled by $0,~2,~4$ ($1,~3,~5,$) on the left (right) panel.~The insets show the corresponding nonlinear SS states.} 
\label{fig:fig3}
\end{figure}
\par
Having discussed the linear properties, we investigate the nonlinearity induced changes when the operating point (angle of incidence) is chosen in the stop gap (marked by a cross in Fig.~\ref{fig:fig2}, i.e., at $\theta=44.81^{\circ}$). Note that the stop gap boundary occurs at $\theta=44.76^\circ$.
An increase in incident intensities leads to an increase (for $\alpha>0$) in the optical widths of the nonlinear layers and hence to bent resonances \cite{ncmt}. This is shown in Fig.~\ref{fig:fig3}, where the scattered intensity is plotted against the incident intensity for both the symmetric (Fig.~\ref{fig:fig3}(a)) and the antisymmetric (Fig.~\ref{fig:fig3}(b)) solutions, respectively. The CPA-like dips are labeled following their linear counterparts in Fig.~\ref{fig:fig2}(b).~The symmetric (antisymmetric) dips $0,~2,~4$ ($1,~3,~5$) occur at $U_i=1.877\times10^{-4},~6.12\times10^{-5},~1.47\times10^{-4}$ ($U_i=5.10\times10^{-5},~9.56\times10^{-5},~2.004\times10^{-4}$). Since, the nonlinearity drives the on-resonant system away from the CPA resonance \cite{ncpa-knr-sdg}, it is difficult to have CPA at both low and high powers for the same system under identical conditions. Till now our attention was on destructive interference and the resulting CPA under optimal conditions. In contrast, one can have constructive interference and super scattering, leading to maximal scattered intensities. Such nonlinear SS states are shown in the insets of  Fig.~\ref{fig:fig3}(a) and Fig.~\ref{fig:fig3}(b).~The lowest order symmetric (antisymmetric) SS states occur at $U_i=4.569\times10^{-3},~6.976\times10^{-3},~1.755\times10^{-2}$ ($U_i=4.872\times10^{-3},~1.136\times10^{-2},~2.555\times10^{-2}$).
 \par
\begin{figure}[]
\center{\includegraphics[width=8cm]{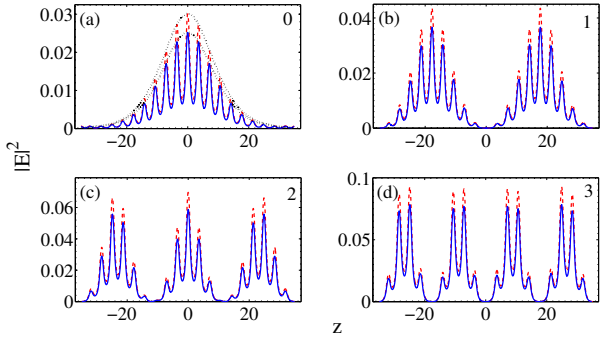}}
\caption{$|E|^2$ (normalized intensity) in the total structure as functions of $z$ corresponding to the nonlinear CPA-like dips. Solid (dashed) curves correspond to exact (approximate) solutions. The results shown in (a)$-$(d) are referred to the points $0-3$ in the Figs. \ref{fig:fig3}(a) and \ref{fig:fig3}(b). The envelops shown by dotted curves in Fig. \ref{fig:fig4}(a) can be fitted with $A/\cosh^2(\beta z)$. }
\label{fig:fig4}
\end{figure}
\begin{figure}[]
\center{\includegraphics[width=8cm]{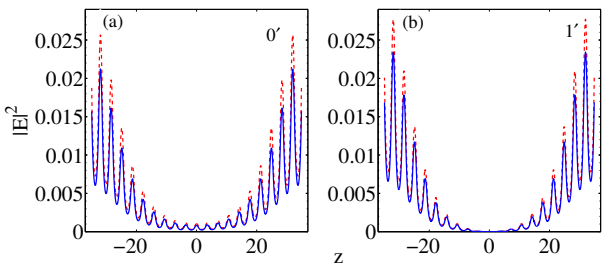}}
\caption{Same as in Fig.~\ref{fig:fig4} but now $|E|^2$ corresponds to nonlinear SS states as labeled by the point  (a) $0^{\prime}$ (symmetric solution), and (b) $1^{\prime}$ (antisymmetric solution) in the insets of  Fig.~\ref{fig:fig3}}.
\label{fig:fig5}
\end{figure}

~Finally we study the normalized intensity $|E|^2$ distribution in the total structure corresponding to the nonlinear CPA-like and SS states by both exact and approximate methods.~The typical field profiles corresponding to nonlinear CPA are shown in Fig.~\ref{fig:fig4}, with left (right) column corresponding to the symmetric (antisymmetric) solutions.~The dashed (solid) curves represent the field profiles as calculated by NCMM (exact integration).~One can see the one-, two-, etc, soliton-like intensity profile  (see Fig.~\ref{fig:fig4}) for CPA-like cases.~For example, the envelope of the distribution in Fig.~\ref{fig:fig4}(a) can be fitted with $A/\cosh^2(\beta z)$ to the exact (approximate) result with $\beta=0.0895$ and $A=0.0252~(0.0301)$ (see Fig.~\ref{fig:fig4}(a) dotted curves). As compared to the CPA states, the SS ones are characterized by an intensity profile with fields localized away from the center (see Fig.~\ref{fig:fig5}(a),~(b) for SS states labeled by $0^{\prime}$ and $1^{\prime}$ in Fig.~\ref{fig:fig3}).~As one can see, both the methods differ very little validating the SVEA in this context with NCMM estimates for $|E|^2$ to be slightly higher than those of the exact results.  
\par
In conclusion, we have studied a Kerr nonlinear periodic layered medium irradiated by $s$-polarized plane waves from both the sides and presented the exact numerical results for CPA. The angle of illumination was chosen inside the stop gap of the linear counterpart. We have shown that CPA in such structures correspond to one- two- etc. soliton-like distributions of the field inside the structure. These results are also obtained by making use of the NCMM offering a true test of the SVEA. The good correspondence of the exact and the approximate results is indicative of the validity of the SVEA and similar approximations of nonlinear optics so far as there are no evanescent waves in the Kerr-nonlinear media. We have also studied the field distributions corresponding to the cases of nonlinear super-scattering (maximal scattering). The excitation of the nonlinear gap solitons with CPA mediated near-null scattering may have interesting application in switching and integrated optical devices.

\end{document}